\documentclass[aps,prl,twocolumn,groupedaddress]{revtex4-2}

\usepackage[hidelinks]{hyperref}
\usepackage{acronym}
\usepackage{graphicx}
\usepackage{amsmath}
\usepackage{upgreek}
\hypersetup{colorlinks=true,linkcolor=blue,citecolor=blue,urlcolor=blue,filecolor=blue}

\begin{document}
\newacro{cdb}[CDB]{coincident Doppler broadening}
\newacro{dbs}[DBS]{Doppler broadening spectroscopy}
\newacro{nepomuc}[NEPOMUC]{neutron induced positron source Munich}
\newacro{frm2}[FRM II]{research neutron source Heinz-Maier Leibnitz}
\newacro{dft}[DFT]{density-functional theory}
\newacro{pas}[PAS]{positron annihilation spectroscopy}

\preprint{}

\title{Breakdown of the Arrhenius Law of the Temperature Dependent Vacancy Concentration in fcc-Lanthanum}

\author{Lucian Mathes} 
\author{Thomas Gigl}
\author{Michael Leitner}
\author{Christoph Hugenschmidt}
\email{christoph.hugenschmidt@frm2.tum.de}
\affiliation{Physik-Department E21 and Heinz Maier-Leibnitz Zentrum (MLZ), Technische Universit\"at M\"unchen, Lichtenbergstra\ss e 1, 85748 Garching, Germany}
\date{\today}

\begin{abstract}
We measured the temperature dependent equilibrium vacancy concentration using in-situ positron annihilation spectroscopy in order to determine the enthalpy $H_\text{f}$ and entropy $S_\text{f}$ of vacancy formation in elementary fcc-La.
The Arrhenius law applied for the data analysis, however, is shown to fail in explaining the unexpected high values for both $S_\text{f}$ and $H_\text{f}$:
in particular $S_\text{f}=17(2)~k_\text{B}$ is one order of magnitude larger compared to other elemental metals, and the experimental value of $H_\text{f}$ is found to be more than three standard deviations off the theoretical one $H_\text{f}=1.46~\text{eV}$ (our \acs{dft} calculation for La at $T=0~\text{K}$).
A consistent explanation is given beyond the classical Arrhenius approach in terms of a temperature dependence of the vacancy formation entropy with $S_\text{f}^\prime=-0.0120(14)~k_\text{B}/\text{K}$ accounting for the anharmonic potential introduced by vacancies.
\end{abstract}

\maketitle

The dominant species of lattice defects, which are thermally created in metal samples, are mono-vacancies.
The enthalpy $H_\text{f}$ and entropy $S_\text{f}$ for vacancy formation in thermal equilibrium are key features for the fundamental understanding of physical processes in crystals such as creation of lattice defects and diffusion properties.

There are several experimental techniques where the measured quantity depends on the concentration of (point) defects. 
Conventionally, measurements of the residual electrical resistivity, which is proportional to the total concentration of all species of lattice defects, are performed to examine the crystal quality or to provide detailed information of defect annealing \cite{Doy62}.  
Differential dilatometry is sensitive to the volume change associated with the formation of lattice defects in the sample and hence allows the estimation of the vacancy concentration in thermal equilibrium but is limited to temperatures close to the melting point \cite{Sie78}. 
In contrast, \ac{pas} is applied as true probing technique to study open-volume crystal defects on an atomic scale.
Due to the efficient trapping in the attractive potential formed by vacancies \cite{Hod70, Sie78} positrons exhibit an outstanding sensitivity for the detection of vacancy concentrations as low as $c_\text{v}\sim10^{-7}$ \cite{Hautojaervi1979}. 

\ac{pas} has been widely applied as non-destructive technique to study the annealing behavior \cite{Man78, Eld76} and the thermal production of vacancies \cite{Ric76}. 
The measurement of the equilibrium vacancy concentrations as a function of temperature in turn allows the determination of the vacancy formation enthalpy $H_\text{f}$.
For a large number of elemental metals and alloys the Arrhenius law has been applied in order to obtain values for $H_\text{f}$ and, to lesser extent, for $S_\text{f}$ (see, e.g., \cite{Ric76, McK72, Mai77, LB91}). 
Besides specific heat measurements on La providing an estimation of $H_\text{f}=1~\text{eV}$ \cite{Aki70}, no further data or detailed studies of the vacancy formation in La is reported to best of our knowledge. 

In this letter we present measurements of the vacancy concentration in thermal equilibrium in fcc-La up to $1020~\text{K}$ using in-situ PAS. 
For comparison of the experimental findings with theory we calculated $H_\text{f}$ for La at $T=0~\text{K}$ based on \ac{dft}. 
For the first data analysis an Arrhenius-like behavior for thermal production of vacancies was assumed, which lead to unexpectedly high values for both the enthalpy $H_\text{f}$ and entropy $S_\text{f}$ for vacancy formation.
This discrepancy was attributed to the fact that the formation entropy is mainly affected by the change of the phonon spectrum of the crystal due to the presence of vacancies. 
In order to obtain a consistent physical explanation we followed a theoretical study by \citet{Gle14} and introduced a temperature dependent vacancy formation entropy.

Lanthanum exhibits phase transitions from dhcp to fcc at $\sim560$~K and from fcc to bcc at $\sim1120$~K \cite{You75}. 
The melting point of La amounts to $1193~\text{K}$ and its density at room temperature is $6.145~\text{g}\,\text{cm}^{-3}$ \cite{Gme86}.
The purity of the sample investigated in the present study is $>99.9\%$.
Since La is highly reactive and would, e.g., oxidize rapidly when exposed to air, it is kept in ethanol during and after sample preparation.
A disc of $4~\text{mm}$ was cut and polished first with SiC grinding paper and subsequently with a $\text{H}_2\text{O}$-free diamond suspension with a final grain size of $1~\upmu\text{m}$.
Possible lattice defects have been annealed by heating the sample up to $1020~\text{K}$ with a heating rate of $13~\text{K}\,\text{min}^{-1}$ and subsequent adiabatic cooling in the \ac{cdb} spectrometer prior the temperature dependent measurements.
Thus the initial $S$ parameter of the as-prepared sample at room temperature was reduced by $4\%$.

Positron-electron annihilation leads predominantly to the emission of two $511~\text{keV}\;\gamma$ quanta.
These photons, which experience a Doppler shift due to the momentum of the annihilating electrons (the momentum of the thermalized positrons is negligible), are examined by \ac{dbs} of the positron annihilation line. 
This broadening strongly depends on the vacancy concentration since the lower annihilation probability of positrons trapped in vacancies with high-momentum core electrons leads to a smaller Doppler-shift compared to annihilation in the pure lattice. 
For the characterization of the Doppler broadening the so-called $S$ parameter is conventionally defined as the fraction of counts in a fixed central region of the annihilation photo peak. 
Hence, compared to the defect-free state the $S$ parameter is usually enhanced for positrons trapped in a vacancy.
For further details of \ac{dbs} we refer to \cite{Coleman2000}.

For the present study, we used the \ac{cdb} spectrometer \cite{Gig17} with a monoenergetic positron beam provided by the \ac{nepomuc} \cite{Hug12} at the \ac{frm2}. 
\ac{dbs} at this spectrometer with an accessible temperature range of $40-1100~\text{K}$ was shown to be particularly suited for the determination of the vacancy concentration, e.g., in Heusler alloys \cite{Hug15a} or the in-situ observation of fast defect annealing after severe plastic deformation \cite{Obe10}. 
In addition, compared to conventional \ac{pas} with $\beta^+$ emitters the usage of a positron beam has the advantage of simple sample heating under ultra high vacuum conditions and that the recorded signal exclusively stems from annihilation events inside the sample (absence of the so-called source component).
The kinetic energy of the positron beam can be varied between $0.1$ and $30~\text{keV}$ and the spot size of the beam at the sample position is typically $250~\upmu\text{m}$.
In this study a maximum implantation energy of $E=28~\text{keV}$ is used corresponding to a mean positron implantation depth of $\bar{z}=1.3~\upmu\text{m}$ in La.
The $S$ parameter is calculated as the fraction of annihilation events of the photo peak in the energy interval $(511\pm1.7)~\text{keV}$.
For the determination of the bulk equilibrium vacancy concentration in La at elevated temperature we performed in-situ \ac{dbs} between $493$ and $1023~\text{K}$ in steps of $10~\text{K}$ using the $28~\text{keV}$ positron beam. 
At each temperature step starting at $1023~\text{K}$ data were recorded for five minutes resulting in about $800\,000$ counts in the $511~\text{keV}$ annihilation photo peak.

\begin{figure}
\includegraphics[width=.9\linewidth]{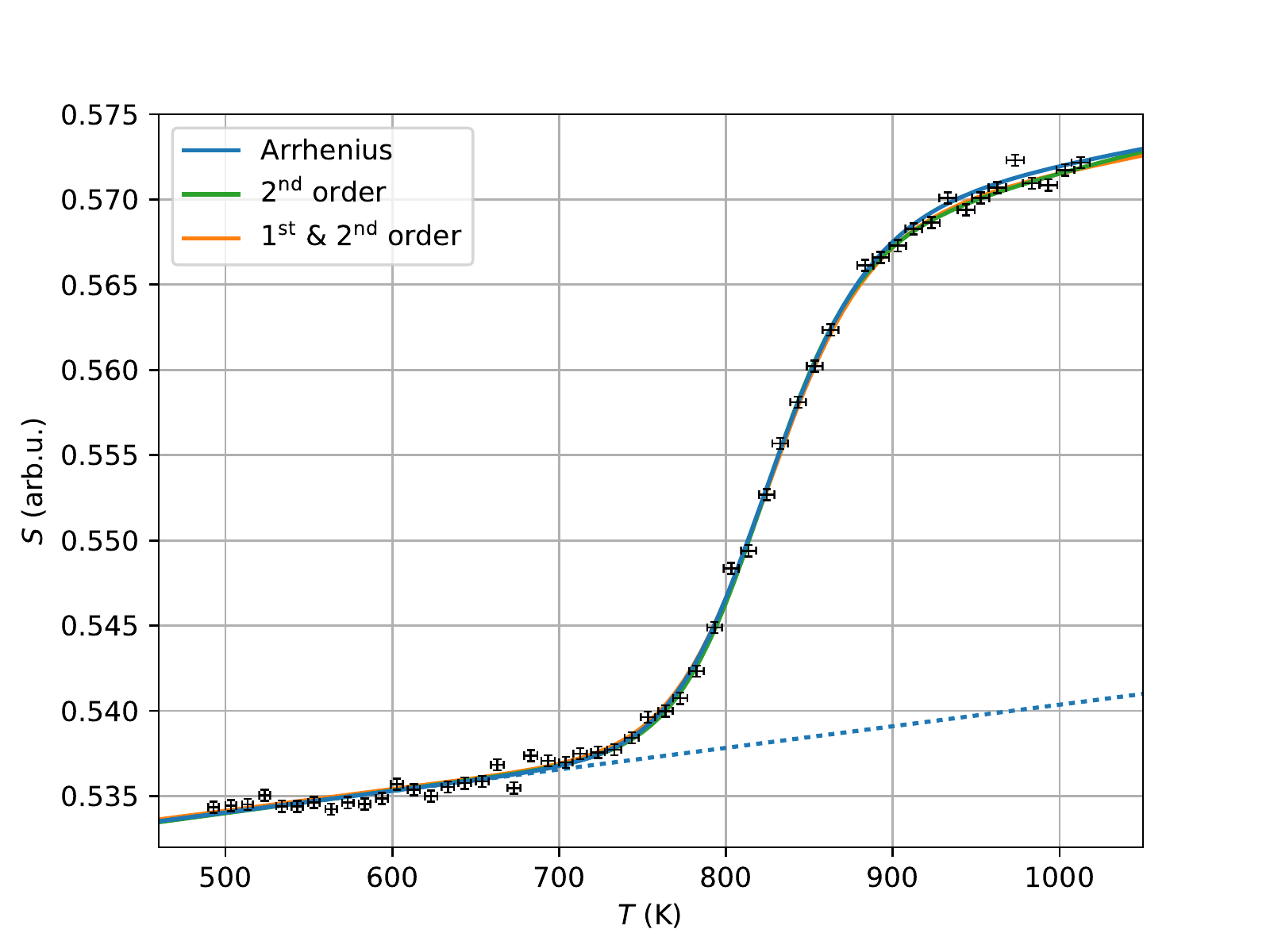}
\caption{Measured $S$ parameter as a function of temperature for La. 
The experimental data (symbols) are fit by a two-state model for positrons annihilating in the bulk or trapped in vacancies according to Eq.~\ref{eq:s} (lines). 
The linear increase of $S$ at lower temperature is well described by the thermal lattice expansion (dashed line).
\label{fig:st} }
\end{figure}

The measured $S$ parameter as function of temperature is shown in Fig.~\ref{fig:st}.
The total increase of $S$ from $493$ to $1023~\text{K}$ amounts to $7\%$.
Up to temperatures of $700~\text{K}$ a linear rise of the $S$ parameter is observed with a slope of $\alpha\approx2.4(3)\times10^{-5}~\text{K}^{-1}$.
This linear increase is attributed to the lattice expansion without significant positron trapping in defects and can be very well explained by the thermal volume expansion coefficient of $\alpha_\text{th,v}=3\alpha_\text{th,l}$, with $\alpha_\text{th,l}\approx8\times10^{-6}~\text{K}^{-1}$ being the thermal linear expansion coefficient of La \cite{Bar57}.
This effect correlates to the decreasing overlap of the positron wave function with those of core-electrons being proportional to the volume expansion of the lattice as observed in studies on, e.g., Al, In, and Pb; the effect of a small contraction of the Fermi surface with higher temperature is negligible \cite{Tri75}.
It is noteworthy that the phase transition for La at $\sim580~\text{K}$ from dhcp to fcc does not affect the linear slope of the $S$ parameter or any other of our fit parameters (within the errors) when starting the fit above $600~\text{K}$. 
This behavior of $S(T)$, however, was expected since the crystal structures fcc and dhcp differ only in the stacking order.
The thermal production of vacancies, which act as efficient positron trapping sites with significantly reduced core annihilation probability, leads to a steeper increase of $S$ above $700~\text{K}$ according to the equilibrium vacancy concentration at the respective temperature.
At about $950~\text{K}$ the $S$ parameter starts to converge due to so-called saturation trapping, since the high vacancy concentration results in trapping of all positrons.

As first proposed by \citet{McK72}, the behavior of the $S$ parameter can be represented by a superposition of two positron states:
positrons annihilate either from a delocalized state in the bulk or from the trapped state in a vacancy with characteristic values $S_\text{b}$ and $S_\text{v}$, respectively.
Hence the $S$ parameter measured at a given temperature $S(T)$ can be described by
\begin{equation}
S(T) = \frac{1}{1 + Q}\,S_\text{b}(T) + \frac{Q}{1 + Q}\,S_\text{v}(T).
\label{eq:s}
\end{equation}
The temperature dependencies of $S_\text{b}$ and $S_\text{v}$ are considered to be linear in $T$ with $(1+\alpha T)$ and $(1+\beta T)$, respectively, and well explain the effect of the lattice expansion as discussed.
The weighting factors in Eq.~\ref{eq:s} are expressed in terms of $Q$ containing properties of the positron and thermodynamical information
\begin{equation}
Q(T) \equiv \frac{S(T) - S_\text{b}(T)}{S_\text{v}(T) - S(T)} = \mu\tau_\text{b}\cdot c_\text{v}(T), 
\label{eq:q}
\end{equation}
with the specific trapping coefficient of a monovacancy $\mu$, the bulk lifetime of a positron $\tau_\text{b}$, and the thermal equilibrium vacancy concentration $c_\text{v}(T)$ at the temperature $T$.
If not explicitly given, $\mu\tau_\text{b}$ has to be estimated to provide a value for $c_\text{v}$ from the measurements. 
Since $\mu\tau_\text{b}$ is actually not known within about one order of magnitude we use the upper limit approximation $\mu\tau_\text{b}\approx4.4\times10^4$ in the following. 
This value is composed of the trapping coefficient $\mu=4\times10^{14}s^{-1}$ (see, e.g., values for Al \cite{Sch87, Wur95} and for Cu, Au, Pt \cite{Sch87b}), and an assumed positron bulk lifetime in La of $\tau_\text{b}\approx110~\text{ps}$, which is in the range of $100~\text{ps}<\tau_\text{b}<120~\text{ps}$ typically obtained for transition metals (see, e.g., calculated values for the fcc metals Cu, Ag, Ni, and Au \cite{Kor96}).

\begin{figure}
\includegraphics[width=.9\linewidth]{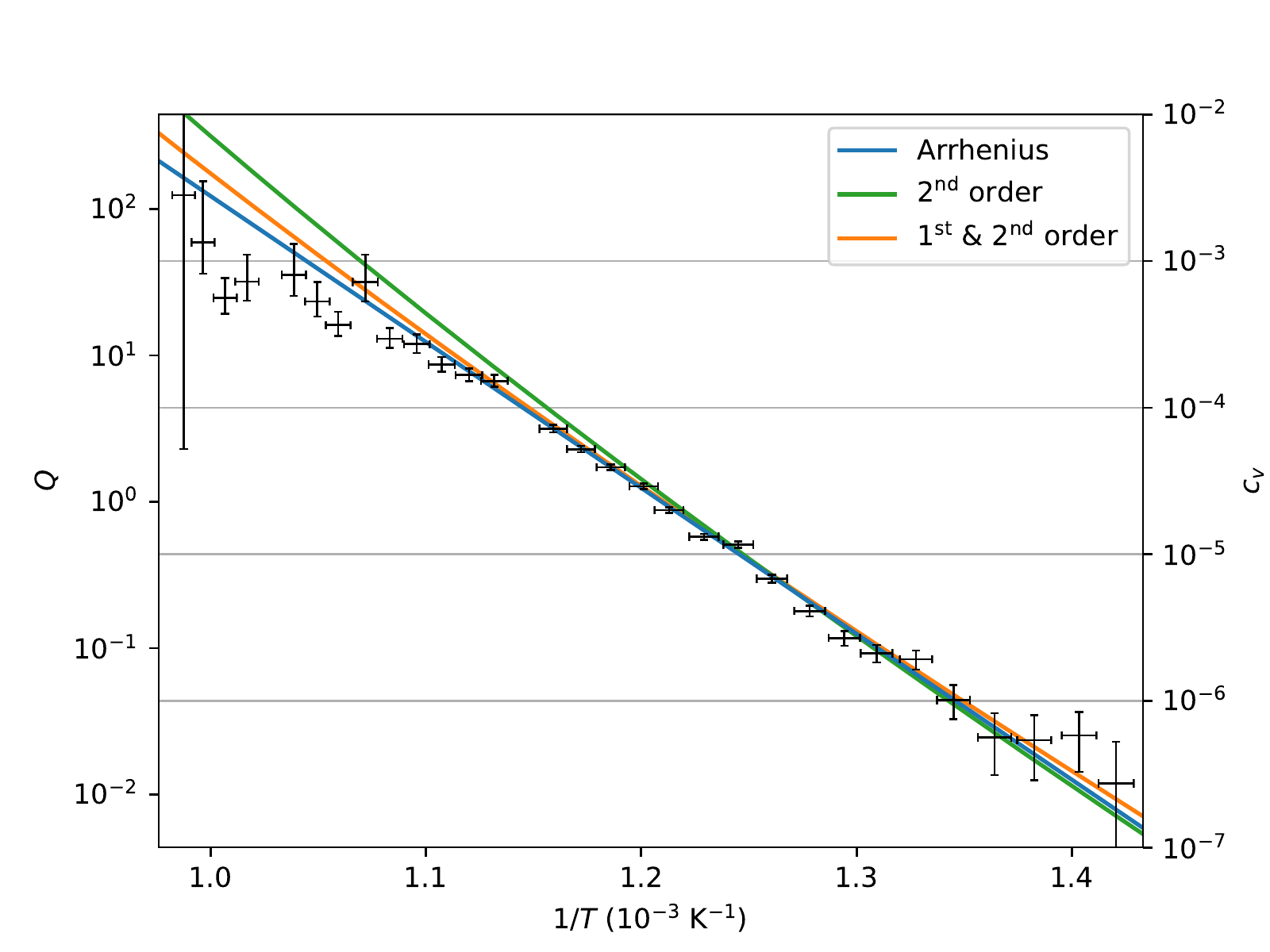}
\caption{Arrhenius plot of the measured data.  
The equilibrium vacancy concentration $c_\text{v}$ (right axis) is deduced from the measured $Q(T)$ using the approximation $\mu\tau_\text{b}\approx4.4\times10^4$ (see Eq.~\ref{eq:q}).
For the different fits (lines) see text.
\label{fig:arrh}}
\end{figure}

Assuming monovacancies being the dominant species of lattice defects and in the limit of non-interacting vacancies their concentration is determined by the Gibbs free enthalpy of vacancy formation $G_\text{f}$ 
\begin{equation}
c_\text{v}(T) = \text{exp}({-G_\text{f}(T)}/{k_\text{B}T}),
\label{eq:cv}
\end{equation}
wherein $k_\text{B}$ is the Boltzmann constant.
The temperature dependence of $G_\text{f}$ is conventionally given by 
\begin{equation}
G_\text{f}(T) = H_\text{f} - TS_\text{f},
\label{eq:gf}
\end{equation}
with enthalpy $H_\text{f}$ and entropy $S_\text{f}$ of vacancy formation, both assumed to be temperature-independent.
It has to be noted that the respective influence of $H_\text{f}$ and $S_\text{f}$ cannot be separated by any experiment measuring the vacancy concentration. 
The data shown in Fig.~\ref{fig:st} were fitted (`Arrhenius', blue line) by applying Eq.~\ref{eq:s} using four parameters ($S_\text{b}^0$, $S_\text{v}^0$, $\alpha$ and $\beta$). 
Tests with different values revealed that any temperature dependence of $\tau_\text{b}$ (see Eq.~\ref{eq:q}) can be well neglected.
By applying Eqs.~\ref{eq:cv} and \ref{eq:gf}  this classical model yields a value for the vacancy formation enthalpy of $H_\text{f}=(1.98\pm0.15)~\text{eV}$.
Fig.~\ref{fig:arrh} shows the recorded data $S(T)$ in the common Arrhenius representation where $H_\text{f}$ is given by the linear slope of the data. 
A linear behavior is observed in the significant region of the covered temperature range. 
Note that the sensitivity threshold for defect spectroscopy with positrons in the order of $10^{-7}$ vacancies per atom is clearly visible and saturation trapping starts around $c_\text{v}=10^{-3}$.

\begin{figure}
\includegraphics[width=.9\linewidth]{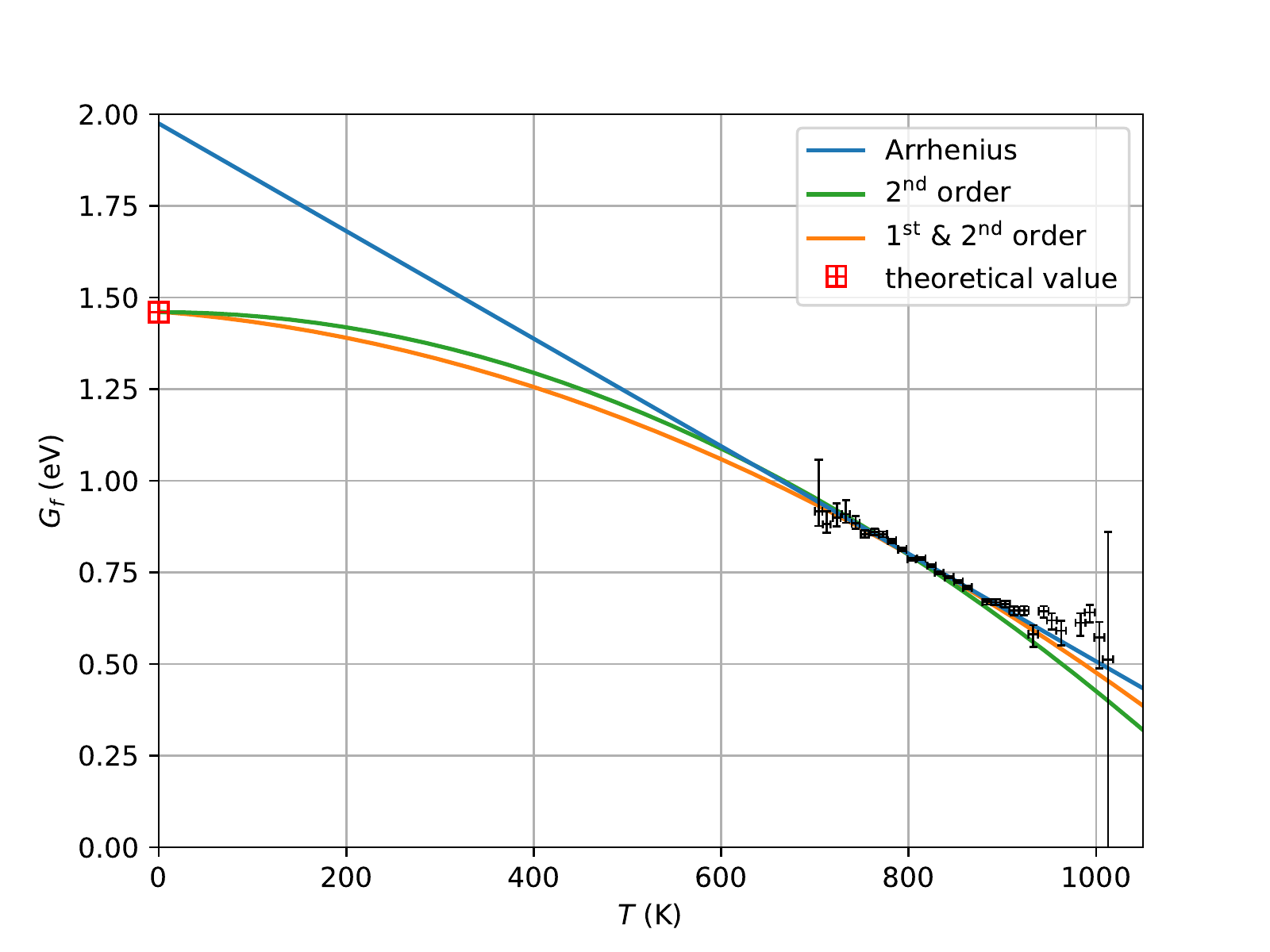}
\caption{Gibbs free enthalpy as function of temperature.
The different curves correspond to Arrhenius-fit (blue line) as well as fits using the Taylor expansion up to the second order with `$1^\text{st}\;\&\;2^\text{nd}\;\text{order}$' (orange line) and without linear term `$2^\text{nd}\;\text{order}$' (green line)
with the calculated value of $G^\text{th}_\text{f}(0)=1.46~\text{eV}$ (red symbol) as boundary condition.
\label{fig:gf}}
\end{figure}

We computed the vacancy formation energy by \ac{dft} with the PBE-generalized gradient approximation \cite{Per96} using the \textsc{abinit} code in the projector-augmented wave framework \cite{Gon09}. 
We used a plane-wave cutoff of $680~\text{eV}$ and a $12\times 12\times 12$ $k$-point grid with respect to the conventional cubic unit cell of the fcc lattice. 
We obtained a lattice constant of $5.29~\text{\AA}$ for the ground state, in perfect agreement with the experimental fcc lattice constant extrapolated to zero temperature \cite{Spe61}. 
The cubic $32-1$-atom supercell with relaxed internal positions but fixed cell dimensions gave a vacancy formation energy of $1.46~\text{eV}$, thus perfectly reproducing the previously reported values of $1.44~\text{eV}$ and $1.46~\text{eV}$ \cite{Ang14,Sha16}.
 To test for a variation of the formation energy with thermal lattice expansion, we performed additional calculations at a lattice constant of $5.32~\text{\AA}$ corresponding to the experimental value around $780~\text{K}$ representative of the temperatures of measurement, which however resulted in only a minute increase of the formation energy to $1.50~\text{eV}$. 
In order to compare experiment with theory we calculated G(T) at each temperature from the measured data by combining Eqs.~\ref{eq:q} and \ref{eq:cv}  and the $S(T)$ fit result (see Fig.~\ref{fig:st}). 
Fig.~\ref{fig:gf} displays the Gibbs free enthalpy as function of temperature with extrapolation of the Arrhenius-fit to $T=0~\text{K}$. 
It becomes obvious that the experimental value $G_\text{f}(0)\equiv H_\text{f}$ is significantly, i.e., more than three standard deviations, off the calculated one. 
For the physical interpretation of the data we formally describe the Gibbs free enthalpy in a more general way by applying the Taylor expansion up to the second order
\begin{equation}
G_\text{f}(T) \approx G(T_0) + G^\prime(T-T_0)+ G^{\prime\prime}(T-T_0)^2/2
\label{eq:qt}
\end{equation}
centered at $T_0=850~\text{K}$, i.e., at the center of the $S(T)$ data set; $G^\prime$ and $G^{\prime\prime}$ are first and second partial derivatives, resepctively, of the Gibbs free enthalpy with respect to temperature. 
Two additional fits with fixed $G^\text{th}_\text{f}(0)=1.46~\text{eV}$ from DFT calculation are performed with and without the linear term (indicated as `$1^\text{st}\;\&\;2^\text{nd}\;\text{order}$' and `$2^\text{nd}\;\text{order}$') and plotted alongside with the Arrhenius-fit in all Figs.~\ref{fig:st} to \ref{fig:gf}. 

The classical Arrhenius law applied to our data for La would yield unrealistic values for both vacancy formation entropy $S_\text{f}$ and Gibbs free enthalpy at $T=0~\text{K}$.
The experimental value of $G_\text{f}(0)$ was found to be $0.52~\text{eV}$ above the calculated one. 
Hence, the failure of the Arrhenius law becomes apparent in Fig.~\ref{fig:gf}: 
consequently, the large difference between experimental and theoretical value of $G_\text{f}(0)$ cannot be explained by the conventionally defined Gibbs free enthalpy being linear in temperature.
Even more importantly, for the vacancy formation entropy we obtain a lower limit of $S_\text{f}=-\partial G/\partial T=(17\pm2)~k_\text{B}$; such high values of $S_\text{f}$ in metals have never been observed to the best of our knowledge. 
It has to be emphasized that this value would correspond to a seven orders of magnitude lower trapping coefficient. 
Typical $S_\text{f}$ for elemental metals, however, are in the range of $0.5-2~k_\text{B}$ \cite{LB91}; i.e., $S_\text{f}\approx1~k_\text{B}$ for fcc and $S_\text{f}\approx2~k_\text{B}$ for bcc crystal lattices \cite{Mai79} and hence about one order of magnitude smaller.  

The influence of possible divacancies has been proven to be negligible in early experiments \cite{Her77} and divacancies, e.g., in Al, were shown to be unstable \cite{Car00}.
Theoretical studies yield that the anharmonicity of lattice vibrations are much more significant than the small effect of possible divacancies \cite{San01}.
This was confirmed in more recent computations of the thermodynamics of divacancies in Al and Cu by \citet{Gle14}, who obtained divacancy concentrations $\leq4\times10^{-3}\cdot c_\text{v}$ even at the melting point.
The effect of positron detrapping from vacancies was found to somewhat influence the measurements at very high temperature as observed for refractory metals such as Ta \cite{Mai77} but is assumed to be negligible in the temperature range of the present study. 
According  to an empiric description based on a temperature dependent vacancy formation enthalpy \citet{Sch90} and assuming a `real' formation entropy of $S_\text{f}=1~k_\text{B}$ for La we would obtain a temperature dependence of $H_\text{f}$ of about $-1.4~\text{meVK}^{-1}$. 
Besides being purely phenomenological this approach relies on a Gibbs free enthalpy depending linearly on temperature, that in turn is unable to explain the theoretically calculated value of $G^{th}_\text{f}(0)$.

In order to describe the exceptionally high $S_\text{f}$ our data can be fitted by using Eq.~\ref{eq:qt} and the calculated value for $G^{th}_\text{f}(0)$.
The resulting best fit (with $G^\prime=1.53\times10^{-3}~\text{eV}/\text{K}$ and $G^{\prime\prime}=-7.8\times10^{-7}~\text{eV}/\text{K}^2$) is displayed as `$1^\text{st}\;\&\;2^\text{nd}\;\text{order}$' (orange line) in the figures.
It has to be emphasized that this rather formal procedure, i.e., the mathematical description of the Taylor expansion of $G(T)$ around $850~\text{K}$, is intrinsically not able to disentangle information of the temperature dependence of $H_\text{f}$ and $S_\text{f}$.
Therefore, we follow the theoretical approach proposed by \citet{Gle14} who performed demanding finite temperature \ac{dft} computations of the Gibbs free energy of vacancy formation by explicitly including anharmonicity due to phonon-phonon interactions, which is of particular importance at high temperatures. 
Compared to the classical Arrhenius behavior deviations for $H_\text{f}$ of $0.15$ and $0.22~\text{eV}$ were found for Al and Cu, respectively \cite{Gle14}.
The formation entropy of vacancies was described to be linear in temperature.
We now apply this physically justified model (local Grüneisen theory) to our experimental results for La and expand $S_\text{f}$ up to the first order in temperature (whereby its constant fraction is neglected as proven to be valid for Al and Cu \cite{Gle14}) 
\begin{equation}
S_\text{f}(T)\approx S_\text{f}^\prime T, 
\end{equation}
where $S_\text{f}^\prime$ is the partial temperature derivative of $S_\text{f}$.
Using Eq.~\ref{eq:gf} we obtain for the Gibbs free enthalpy
\begin{equation}
G_\text{f}(T)=H_\text{f}^\text{0K}-T^2S_\text{f}^\prime.
\end{equation}
Fitting the data hence requires only two free parameters $H_\text{f}^\text{0K}$ and $S_\text{f}^\prime$.
According to our \ac{dft} calculation the first one is found to be $H_\text{f}^\text{0K}=1.46~\text{eV}$ and the second one is $S_\text{f}^\prime=-0.0060(14)~k_\text{B}/\text{K}$.  
The according fit depicted as `$2^\text{nd}\;\text{order}$' is shown in Figs.~\ref{fig:st} to \ref{fig:gf}.
 Using this model the linear increase with $\alpha\approx2.5(2)\times10^{-5}~\text{K}^{-1}$  in the temperature range $480$ to $700~\text{K}$ is indistinguishable from the other fits as shown in Fig.~\ref{fig:st} but clearly deviates from the Arrhenius law at lower temperature (see Fig.~\ref{fig:gf}).

In summary we found unexpectedly high discrepancies for both $H_\text{f}$ and $S_\text{f}$ by applying the classical Arrhenius interpretation to our data obtained by in-situ \ac{pas} at high temperatures:
the Gibbs free enthalpy at $T=0~\text{K}$ was more than three standard deviations higher than that resulting from our \ac{dft} calculation.
Even more surprising, however, is the exceptional high value for the entropy $S_\text{f}$, which  was found to be about one order of magnitude higher than typical ones for elemental metal crystals.
In this letter, a consistent explanation is given in terms of a temperature dependent vacancy formation entropy taking into account the anharmonicity of phonons introduced by the presence of monovacancies in the crystal lattice.

\end{document}